\documentclass[prb, twocolumn]{revtex4}
\usepackage[english]{babel}
\usepackage[dvips]{graphicx}

\begin{document}
\title{Electromechanical instability in suspended carbon nanotubes}
\author{L. M. Jonsson} \email{mjonsson@fy.chalmers.se}
\affiliation{Department of Applied Physics, Chalmers University of
Technology, SE - 412 96 G{\"o}teborg,
Sweden}
\author{L. Y. Gorelik} \affiliation{Department of Applied Physics,
Chalmers University of Technology, SE - 412 96 G{\"o}teborg, Sweden}
\author{R. I. Shekhter} \affiliation{Department of Physics,
G{\"o}teborg University, SE - 412 96 G{\"o}teborg, Sweden}
\author{M. Jonson} \affiliation{Department of Physics,
G{\"o}teborg University, SE - 412 96 G{\"o}teborg, Sweden}
%\keywords{NEMS, Electroemchanical instability, CNT}
\begin{abstract}
We have theoretically investigated electromechanical properties of
freely suspended carbon nanotubes when a current is injected into the
tubes using a scanning tunneling microscope. We show that a
shuttle-like electromechanical instability can occur if the bias
voltage exceeds a dissipation-dependent threshold value. An
instability results in large amplitude vibrations of the carbon
nanotube bending mode, which modify the current-voltage
characteristics of the system.
\end{abstract}

\maketitle Nanoelectromechanical systems (NEMS) have been a subject of
extensive research in recent years. The possibility of combining
electrical and mechanical degrees of freedom on the nanoscale may give
rise to technological advantages as well as manifestations of
fundamental physical phenomena. From a technological point of view the
interest is largely due to the many applications that may be realized
using NEMS. \cite{Cleland} Among the many NEMS phenomena of
considerable physical interest, we focus in this Letter on the
``shuttle" instability that is expected to occur in a specific class
of systems. This phenomenon was predicted by Gorelik \emph{et al.}
\cite{Gorelik} and have been theoretically investigated in a
number of publications, see Shekther \emph{et al.} \cite{ShuttleReview1} 
for a review.  The typical shuttle system consists of a movable
conducting grain placed in an elastic medium between two metallic
leads. When the system is DC-biased over a certain threshold, an
electromechanical instability occurs, resulting in large-amplitude
mechanical vibrations and a shuttle current proportional to the
vibration frequency. Different experimental approaches have been used
to build and investigate various versions of this
system. \cite{Scheible, Park} However, some problems persist and a
true shuttle instability has not been observed. Carbon nanotubes (CNT)
are materials that have proved to be strong candidates for building
NEMS. The electrical properties of CNTs depend on their detailed
structure and the mechanical properties are extraordinary. A very high
Young's modulus and low mass give typical resonance frequencies in the
GHz-regime and the tubes can be mechanically deformed without
breaking. Progress in fabrication of CNTs shows that it is
possible to build structures, for example suspended tubes \cite{Lee,
LeRoy, Sazonova} and nanorelays, \cite{LeeRelay} where CNT mechanical
degrees of freedom are important.
The work presented here was done in response to recent experimental
progress by LeRoy \emph{et al.}  \cite{LeRoy} in measurements on
suspended carbon nanotubes (CNT) using a scanning tunneling microscope
(STM). Peaks in the $dI/dV-V$ characteristics when current is injected
into the free hanging part of the nanotube are attributed to inelastic
tunneling of electrons that interact with the radial breathing
mode (RBM) of the tube. This is a manifestation of a coupling between
the electronic and mechanical degrees of freedom in the system,
allowing for additional channels of charge transport through the
system by means of vibronic assisted tunneling. We would like to point
out that in addition a different electromechanical effect --- an
electromechanical shuttle-like instability --- may occur in the
system. Such an instability occurs when the ground state is unstable
with respect to small spatial deviations of the nanotube from its
equilibrium position.  Small displacements grow in time and develop
into limit cycle oscillations that modify the transport properties.
The shuttle instability is fundamentally different from vibronic
assisted tunneling, in the sense that it results in a condensate of
coherent phonons \cite{Fedorets2} and hence
large-amplitude mechanical bending mode vibrations.  Below we will
show that a shuttle-like instability may indeed occur in the suspended
CNT-structure. It may therefore be a useful experimental system for
investigating shuttling phenomena, also in the quantum regime.
Quantum effects are expected to be important if the amplitude $\Delta
X_0$ of the zero-point bending-mode oscillations is at least as large
as the tunneling length $\lambda \sim 0.5$ \AA\, -- a length that
characterizes the tunnel barrier separating the STM tip and the
nanotube.  Theoretically, the shuttle system has been modeled in both
the quasi-classical \cite{Fedorets} and the quantum mechanical limits,
\cite{Fedorets2, Novotny} while so far the quantum regime has not been
experimentally accessible.  We will show below that the suspended CNT
structure is the best candidate system for experiments that probe the
quantum regime of shuttling.

\begin{figure}
\includegraphics[width=7cm]{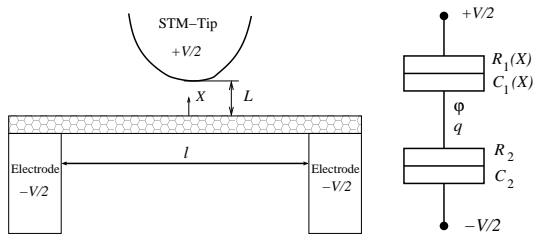}\hspace{0.1cm}
\caption{Model system (left) and equivalent circuit (right). A
carbon nanotube of length $l$ is suspended over a trench; the tube
ends are connected to metal electrodes. The tube can be
mechanically deformed and $X$ measures the deflection of the
center of the tube from its equilibrium position. An STM-tip is
positioned a distance $L$ above the center of the tube and a
DC-bias voltage $V$ is applied between the STM and the electrodes.
The CNT is treated as a metallic island with excess charge $q$ and
electrostatic potential $\varphi$. The STM-tube contact is a
tunnel junction characterized by resistance $R_1(X)$ and
capacitance $C_2 (X)$. The contacts between tube and electrodes
are also tunnel junctions, characterized by a constant resistance
$R_2$ and a constant capacitance $C_2$.} \label{fig:modelsystem}
\end{figure}

In the present work we investigate a system where an STM-tip is used
to measure the current through a suspended CNT as in the experiment by
LeRoy \emph{et al.}  \cite{LeRoy} We use the model system depicted in
Figure \ref{fig:modelsystem} and consider a metallic CNT (single or
multi-wall) suspended over a trench. The nanotube is treated as a
classical continuum beam with length $l$, inner radius $r_{{\rm i}}$
and outer radius $r_{{\rm o}}$. An STM-tip is positioned near the
center of the trench in close proximity to the nanotube; the distance
from the undeformed tube to the tip is \mbox{$L \sim 1$ nm}. The STM
acts as an electrode, which is connected to the tube by a tunnel
junction. The STM-tube separation is not fixed but depends on the
deflection $X$ of the tube from the equilibrium straight position.
Both the resistance $R_1$ and the capacitance $C_1$ characterizing
this junction are functions of this parameter.  The ends of the tube
rest on electrodes to which they are electrically connected via tunnel
junctions rather than ohmic contacts -- the most likely scenario for
CNT-metal junctions. The parameters for these junctions can be
described by a single constant capacitance $C_2$ and a single constant
resistance $R_2$. A DC-bias voltage $V$ is applied between the STM and
the electrodes. The time scale for charge redistribution in the tube
is fast compared to other time scales in the system allowing one to
treat the nanotube as a metallic island with an excess charge $q$ and
electrostatic potential $\varphi$. A capacitive force $F_{{\rm cap}}$
that depends on the potential drop between the STM and the CNT
attracts the tube toward the STM-tip.

We first discuss the possibility for an electromechanical instability
to occur in our model system.  For this purpose we restrict 
ourselves to the case when $R_1 \sim R_2$ in equilibrium. Because
capacitances are not strongly dependent on geometry, we furthermore
neglect the variation of the capacitance with position. Since $R_1$
depends exponentially on $X$, $R_2 \gg R_1$ when the tube-STM
separation decreases significantly on the scale of the tunneling
length and $R_1 \gg R_2$ when the separation increases
significantly. This implies that when the tube is close to the
STM-tip, the time scale $\tau_1$ for charge exchange between the STM
and the tube is much shorter than the time scale $\tau_2$ for charge
exchange between the tube and the electrodes. Hence, the potential on
the tube approaches the STM-potential. In the opposite limit, when the
tube is far from the STM, $\tau_1 \gg \tau_2$ the tube potential 
approaches the potential of the electrodes. Consider now oscillations
with frequency $\omega$ so that the tube position alternates between
these limiting values. Finite values of the parameters $\omega
\tau_{1,2}$ lead to retardation, which implies that the potential on
the tube does not have time to reach its equilibrium value determined
solely by the tube position. Rather, the potential depends on the tube
position at earlier times as well, \emph{i.e.}  on the tube velocity
and exceeds the equilibrium value when the tube moves away from
the tube and vice versa.  It follows that the force on the tube is
correlated with the tube velocity, $\overline{\dot X F_{{\rm cap}}}
\ne 0$, and that mechanical work will be performed on the tube during
the oscillatory motion.

These qualitative arguments show that the necessary force-velocity
correlation for an instability to occur will develop. However, the
performed work must be positive and large enough to exceed the
energy that can be dissipated. If this condition is met, a small
deviation from the equilibrium position of the tube grows in
time and results in limit cycle oscillations. The instability will
change the current flow through the system, the nature of this
change will be discussed below.

To support the qualitative arguments, we have made quantitative
calculations for our model, focussing on the fundamental bending mode
of the doubly clamped nanotube. The eigenfrequency of this mode can be
obtained from linear elasticity theory as $\omega \approx
(22.4/l^2)\sqrt{EI/\rho S}$, where $E$ is the Young's modulus of the
CNT, typically of order\cite{Wong, Treacy} 1 TPa , $I = \frac{\pi}{4}
(r_{{\rm o }}^4-r_{{\rm i }}^4)$ is the area moment of inertia of the
cross section, $\rho$ is the mass density and $S$ is the tube
cross-section area. Excitation of the fundamental bending mode can be
described by a single parameter $X(t)$, which denotes the deflection
of the center of the tube from its equilibrium position. In the limit
of small deflections ($X \ll l$) the potential energy of the bent tube
can be expressed as $U = kX^2/2$, where $k$ is an effective spring
constant for the tube. Hence, the equation of motion for the
mechanical degree of freedom is
\begin{equation}
m_{{\rm eff}} \ddot X + \gamma \dot X + k X = F_{{\rm cap}}.
\label{eq:mech}
\end{equation}
The effective mass of the tube $m_{{\rm eff}} = k/\omega^2 \approx
0.4 m_{{\rm tot}}$ in eq~\ref{eq:mech} is proportional to the
total tube mass $m_{{\rm tot}}$ and a phenomenological viscous
damping $\gamma \dot X$ characterized by a quality factor $Q =
\omega m_{{\rm eff}} /\gamma $ has been introduced. To evaluate
the capacitive force we use model capacitances for the junctions.
The STM-tube junction is approximated by a parallel plate
capacitance $C_1 = C_0/(1-X/L)$, where $C_0$ is the typical
capacitance between the STM and the tube and $L$ is the
equilibrium STM-tube separation. The tube-electrodes capacitance
$C_2$ is a constant.

It is convenient to work with the excess charge $q$ on the tube as
a variable instead of the tube potential.  The capacitive force
then takes the form \mbox{$F_{{\rm cap}} = F_0 (C_2 V -
q)^2/(1-\alpha X)^2$}, where $F_0$ depends on the capacitances and
$\alpha \sim 1/L$. Charge is treated as a continuum variable,
which is valid if the charge exchange rate is high and the
temperature not too low. Using an exponential form for the
STM-tube resistance, \mbox{$R_1(X) = R_0 \exp (-X/\lambda)$}, we
derive a differential equation for the time evolution of the
charge according to
\begin{equation}
\dot q = \frac{V}{2} \bigg( G_- - \frac{C_-}{C_\Sigma} G_\Sigma
\bigg) - q \frac{G_\Sigma}{C_\Sigma}\,, \label{eq:charge}
\end{equation}
where $C_\Sigma(X) = C_1(X) + C_2$ and $C_- = C_1(X) - C_2$ are
the sum and difference of the junction capacitances and
$G_\Sigma(X) = 1/R_1(X) + 1/R_2$ and $G_- = 1/R_1(X) - 1/R_2$ are
the sum and difference of their conductances.

Equations~\ref{eq:mech} and \ref{eq:charge} define our model
system, which we will now analyze both analytically and
numerically. Defining $(X_0, q_0)$ as their steady state solution
we linearize these equations with respect to small deviations from
the steady state. We find that a deviation grows with time for
fixed $V$ if the dissipation is below a threshold value
$\gamma_{{\rm thr}}$. This implies that a shuttle-like instability
does occur in the system. However, the threshold dissipation is
not a positive definite function of the system parameters and if
it is negative no instability will occur. In this case the
force-velocity correlation leads to intrinsic dissipation in the
system, which suppresses any mechanical deviation and prevents an
instability from developing. This effect originates from the
position dependence of the STM-tube capacitance, which contributes
a term to the force-velocity correlation that has the opposite
sign compared to the term coming from the position dependence of
the resistance. Since capacitances are not strongly dependent on
geometry ($L \gg \lambda$), the correlations due to the change in
resistance usually dominates. Only in the limit $R_2 \gg R_0$ does
the position dependence of the capacitance become important since
the potential of the tube is then close to the STM-potential and
the voltage drop over the STM-tube junction does not change
appreciably with tube position.

We can investigate the threshold dissipation analytically in
various limits. We find that the threshold dissipation goes to
zero in the limits $(R_{0} \rightarrow 0,R_{2} \rightarrow 0)$,
$(R_{0} \rightarrow \infty, R_{2} \rightarrow \infty)$ and $R_{0}
\rightarrow \infty$ while it approaches a negative value if $R_{2}
\rightarrow \infty$. First consider the limit $(R_{0} \rightarrow
0, R_{2} \rightarrow 0)$ which implies $(\omega
\tau_{1}\rightarrow 0 , \omega \tau_{2}\rightarrow 0)$ and a
sufficiently fast charge exchange rate for the tube potential to
always have its equilibrium value. Retardation is negligible and
no mechanical work is performed on the tube. In the limit $(R_{0}
\rightarrow \infty, R_{2} \rightarrow \infty)$, $(\omega \tau_{1}
\rightarrow \infty, \omega \tau_{2} \rightarrow \infty)$ charge
exchange is a slow process and the tube potential is approximately
constant during tube oscillations, making retardation a small
effect also in this limit.  Finally, in the limits
$R_{0}\rightarrow \infty$ or $R_{2}\rightarrow \infty$, we have
$\omega \tau_2 \gg \omega \tau_1$ or $\omega \tau_2 \gg \omega
\tau_1$, respectively, and the potential on the tube is mainly
determined by the coupling to the electrodes or the STM. The
latter case leads to dissipation as discussed previously and the
former leads to a small force-velocity correlation since
corrections to the equilibrium potential is small.

\begin{figure}
\includegraphics[width=7cm,height=5cm]{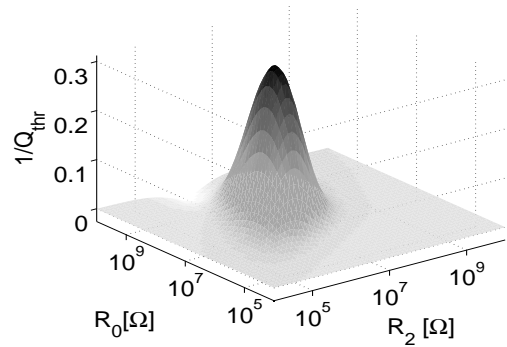}
\caption{Threshold dissipation as a function of junction
resistances for \mbox{$V = 1$ V}. Dissipation is proportional to
$1/Q$, the inverse quality factor for mechanical oscillations, and
an instability occurs if $1/Q < 1/Q_{{\rm thr}}$.  Large values of
$1/Q_{{\rm thr}}$ correspond to optimal conditions for an
instability to occur, and are found when the junction resistances
are of the same order of magnitude. In the limit of highly
asymmetric junctions $1/Q_{{\rm thr}}$ is small.} \label{fig:Qthr}
\end{figure}

Our qualitative analysis suggests that optimal conditions for an
instability can be found in a symmetric setup where $R_{0} \sim
R_{2}$.  The threshold dissipation for a typical set of parameters
\footnote{$C_0 = C_2 = 5 \cdot 10^{-18}$ F, $l = 200$ nm, $r_{\rm o} =
5$ nm, and $r_{\rm i} = 1 $ nm.} is plotted as a function of junction
resistances in Figure \ref{fig:Qthr}, where a region of optimal
conditions for the instability can be identified. This region
corresponds to a symmetric case, where $\omega \tau_1 \sim \omega
\tau_2 \sim 1$, in support of the qualitative conclusion that
resistances of the same order of magnitude are optimal. Because of
this, we do expect poor conditions for an instability in the
experiment by LeRoy \emph{et al.}  \cite{LeRoy} since their junction
resistances are highly asymmetric.

The above arguments show that the occurrence of a shuttle-like
electromechanical instability for a fixed voltage crucially 
depends on the Q-factor of the bending mode vibrations and the
junction parameters. Experimental Q-values of the order of
$10^1-10^3$ have been reported for CNTs, \cite{Sazonova,
Poncharal} corresponding to weak dissipation and an instability
feasible with bias voltages smaller than 1 V for a large range of
parameters.

\begin{figure}
\includegraphics[width=5.2cm, height=3.4cm]{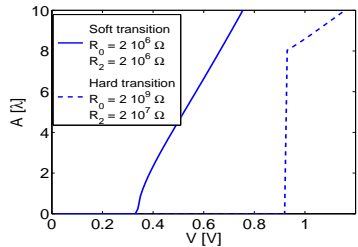}
\caption{An electromechanical instability occurs when the
bias voltage exceeds a threshold value, which results in
bending mode vibrations of the nanotube with amplitude $A$. The
amplitude either increases smoothly (solid line) or in a step-like
fashion (dashed line) as a function of voltage.  Here $Q = 50$.}
\label{fig:1}
\end{figure}

\begin{figure}
\includegraphics[width=6.4cm, height=3.3cm]{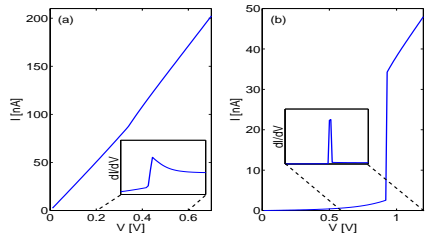}
\caption{$I-V$ characteristics corresponding to the soft (a) and
the hard (b) transition in Figure \ref{fig:1}. Insets depict the
$dI/dV-V$ characteristics in the vicinity of the threshold
voltage. The soft transition manifests itself as a jump in
the value of $dI/dV$, whereas the hard transition gives rise to a
peak in the $dI/dV-V$-curve.} \label{fig:spectroscopy}
\end{figure}

In order to investigate the instability in more detail we solve
the system of equations \ref{eq:mech} and \ref{eq:charge}
numerically for different bias voltages using a typical set of
parameters.\footnotemark[\value{footnote}] The resuts confirm the
analytical conclusions, i.e. if the voltage exceeds a
certain threshold an electromechanical instability occurs
unless the threshold dissipation is negative. The amplitude of the
limit cycle mechanical vibrations is shown as a function of bias
voltage in Figure \ref{fig:1}, where a threshold is clearly
visible when the vibration amplitude $A$ becomes non-zero. The
instability is associated with different characteristics depending
on the type of transition. The vibration amplitude is either a
step function of voltage (hard excitation) or a smooth function
(soft excitation). In the case of a hard excitation the current
will be a step-like function of voltage, whereas a soft excitation
leads to a jump in the first derivative of the current. This can
be seen in Figure \ref{fig:spectroscopy}, where the $I-V$ and the
$dI/dV-V$ curves corresponding to the instabilities in Figure
\ref{fig:1} are shown.


As we have seen, the electromechanical instability changes the
transport properties of the system, leading to a step in the $I-V$
or $dI/dV-V$ characteristics depending on the type of transition
that occurs. These are features that may be used to detect the
instability. Furthermore, theoretical studies of the current noise
in the ordinary shuttle \cite{Gorelik}  have shown that the
instability is associated with an increase of current noise.
\cite{IsacssonNordNoise} Since we expect a shuttle-like
instability in the CNT system to result in a qualitatively similar
noise increase, noise measurements may also be used to identify
the instability. A third approach to detecting the instability is
to apply a small modulation to the bias voltage in analogy with
the Sazonova \emph{et al.} mixing experiment.\cite{Sazonova}

Quantum effects in shuttling have been considered
theoretically \cite{Fedorets2, Novotny}, but they have so
far not been accessible by experiments. With the suspended CNT
system one can control the amplitude of the zero point
oscillations by changing the length of the tube. It should
therefore be possible to experimentally probe the regime where
quantum effects are important by using longer tubes. An estimate
of the ratio $2\Delta X_0 /\lambda$ --- of order one or larger in
the quantum regime --- for different parameters shows it to be in
the range $(0.5-1)$ $(l /\mu {\rm m})^{1/2}$ for a single wall
carbon nanotube. This is a larger value than can be estimated for
other realizations of the shuttle \emph{e.g.} one finds $2\Delta
X_0 /\lambda \sim 0.2-0.3$ in the C$_{60}$-SET device used by Park
\emph{et al.} \cite{Park} The suspended CNT-shuttle is so far the
system where quantum corrections will be most significant.

The typical STM-tube separation is of order 1~nm and, consequently,
surface forces between the tube and the STM will be
important. However, the influence of these forces is mainly
quantitative, it does not change the qualitative conclusion that an
instability occurs. The scale of these forces for interactions between
a CNT and a metallic electrode is \cite{Jonsson} of the order of a few
nN. In the suspended CNT-structure the scale of the elastic forces can
be changed by orders of magnitude by varying the tube length and
thickness. Consequently, the importance of the surface forces relative
to the elastic forces depends on the tube geometry.

To conclude, we propose that a shuttle-like electromechanical
instability can be experimentally studied by using suspended
carbon nanotubes in tunneling contact with supporting metal
electrodes and an STM tip. We support this claim with an analysis
of a classical model of such a system and seek the optimal
conditions for an instability to occur. The shuttle-like
instability is most likely to occur in the symmetric case, when
all the tunneling resistances are of the same order of magnitude.
We demonstrate that the effect of the instability is to change the
$I-V$ characteristics and, finally, we show that the suspended
nanotube system is the first, where the quantum regime may be
approached experimentally.

{\bf Acknowledgment.} Financial support from the Swedish Research
Council and the Swedish Foundation for Strategic Research is
gratefully acknowledged. This work was in part supported by EC FP6 funding
(contract no. FP6-2004-IST-003673, CANEL). This publication reflects
the views of the authors and not necessarily those of the EC. The
Community is not liable for any use that may be made of the
information contained herein.

\end{document}